\documentclass[journal]{IEEEtran}
\usepackage{graphicx,amssymb,amsmath}
\usepackage[noadjust]{cite}%cite
\usepackage{setspace}               % set line space
\usepackage{stfloats}

\usepackage{amsthm}
\usepackage{amsmath}
\usepackage{flushend}
\usepackage{cases,subeqnarray}
\usepackage{bm,multirow,bigstrut}
\usepackage{textcomp}
\usepackage{latexsym,bm}
\usepackage{booktabs,changebar}
\usepackage{xcolor}
\usepackage{colortbl}
\usepackage{mathtools}
\usepackage{dsfont}
\usepackage{extarrows}
\usepackage{mathrsfs}
\usepackage{cite}
\usepackage{bm}
\usepackage{cleveref}
\usepackage{multicol}       % table
\usepackage{multirow}       % table
\usepackage{array}          % table
\definecolor{crimson}{RGB}{192,0,0}         % color crimson
\definecolor{navy}{RGB}{47,85,151}         % color crimson

\makeatletter                               % algorithm
\newif\if@restonecol
\makeatother

\makeatletter
\newif\if@restonecol
\makeatother

\usepackage[linesnumbered,ruled,vlined]{algorithm2e}
\usepackage{algpseudocode}
\usepackage{amsmath}
\renewcommand{\arraystretch}{1.5} %

\theoremstyle{plain}

\newtheorem{lemma}{Lemma}

\theoremstyle{plain}

\newcommand{\argmin}[1]{{\underset{{#1}}{\mathrm{arg\,min}}}}
\newcommand{\vect}[1]{\mathbf{#1}}

\def\diag{\mathrm{diag}}

\def\Htran{\mbox{\tiny $\mathrm{H}$}}
\def\Ttran{\mbox{\tiny $\mathrm{T}$}}
\def\CN{\mathcal{N}_{\mathbb{C}}} %Complex Gaussian
\begin{document}

%----------------------------title&author&thanks----------------------------
\title{Sparse Large-Scale Fading Decoding in \\ Cell-Free Massive MIMO Systems}
\author{\vspace{-1.0em}\setlength{\baselineskip}{11pt}
Shuaifei~Chen$^{\ast, \star}$, Jiayi~Zhang$^{\ast, \star}$, Emil Bj{\"o}rnson$^\dag$, {\"O}zlem Tu{\u{g}}fe Demir$^\dag$, and Bo~Ai$^{\star,\ddag}$\\
{\small $^\ast$School of Electronic and Information Engineering, Beijing Jiaotong University, Beijing 100044, China}\\
{\small $^\star$Frontiers Science Center for Smart High-speed Railway System, Beijing Jiaotong University, Beijing 100044, China}\\
{\small $^\dag$Department of Computer Science, KTH Royal Institute of Technology, SE-16440 Kista, Sweden}\\
{\small $^\ddag$State Key Laboratory of Rail Traffic Control and Safety, Beijing Jiaotong University, Beijing 100044, China}\\
{\small Emails: \{shuaifeichen, jiayizhang, boai\}@bjtu.edu.cn, \{emilbjo, ozlemtd\}@kth.se}\\
\vspace{-2.75em}
\thanks{This work was supported in part by the Fundamental Research Funds for the Central Universities under Grant 2021YJS001 and in part by the FFL18-0277 grant from the Swedish Foundation for Strategic Research.}
}

\pagestyle{empty}       % no page number for the second and the later pages
\maketitle
\thispagestyle{empty}   % no page number for the first page
%----------------------------abstract----------------------------
\begin{abstract}
Cell-free massive multiple-input multiple-output (CF mMIMO) systems are characterized by having many more access points (APs) than user equipments (UEs).
A key challenge is to determine which APs should serve which UEs.
Previous work has tackled this combinatorial problem heuristically.
This paper proposes a sparse large-scale fading decoding (LSFD) design for CF mMIMO to jointly optimize the association and LSFD.
We formulate a group sparsity problem and then solve it using a proximal algorithm with block-coordinate descent.
Numerical results show that sparse LSFD achieves almost the same spectral efficiency as optimal LSFD, thus achieving a higher energy efficiency since the processing and signaling are reduced.
\vspace{-0.5em}
\end{abstract}

%----------------------------keywords----------------------------
\vspace{-1.2em}

%\newpage

\section{Introduction}\vspace{-0.2em}

The sixth-generation (6G) communications are expected to achieve $100 \times$ spectral efficiency (SE) and energy efficiency (EE) gains over the fifth-generation networks \cite{saad2019vision}.
This requires a denser network infrastructure and cell-free (CF) operation that shifts the communication network from cell-centric to user-centric, to provide ubiquitous coverage, improved network throughput, and reduced energy consumption \cite{zhang2019multiple,chen2021survey,demir2022cell}.

\begin{figure}[t!]
\vspace{0.5em}
\centering
\includegraphics[scale=1]{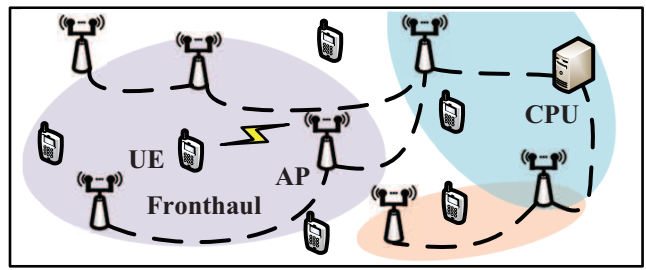}\vspace{-0.5em}
\caption{An illustration of the considered CF mMIMO systems.
\label{fig:system}}
\vspace{-1.75em}
\end{figure}

User-centric CF massive multiple-input multiple-output (mMIMO) is a promising paradigm for increasing the average and worst-case data rates in 6G networks \cite{cellfreebook}.
In CF mMIMO, a large number of distributed access points (APs), coordinated by a central processing unit (CPU), serve the user equipments (UEs) by coherent joint transmission and reception, as illustrated in Fig.~\ref{fig:system}.
Due to the UE-AP-CPU architecture, the signal processing tasks can be distributed between the APs and the CPU in different ways \cite{nayebi2016performance,bjornson2019making}.
Especially, \cite{nayebi2016performance} applied large-scale fading decoding (LSFD), which is a two-stage decoding procedure where \emph{all} APs first compute soft estimates of the data signal from {\it all} UEs and then deliver them to the CPU for final decoding.
Although this approach achieves a good compromise between SE and computational complexity compared to fully centralized processing \cite{cellfreebook}, it might not be energy efficient in a large network in its original form.
It is unnecessary for an AP to waste power and computational resources on UEs with weak channels to it \cite{buzzi2019user}. The geometry induces a sparse structure on the practically meaningful AP-UE associations.
Prior work has suggested associating each UE with a subset of APs (typically in a heuristic non-SE-optimizing manner \cite{bjornson2020scalable}) in advance and then excluding APs not serving this UE when computing the LSFD vector \cite{nayebi2016performance,bjornson2020scalable}.
However, treating the AP-UE association as a separate combinatorial problem from the LSFD design, which is used to maximize the SE \cite{nayebi2016performance}, is suboptimal.
That motivates us to take the association as a part of the LSFD design and exploit sparsity  methods to jointly solve the association problem.

Sparse optimization has attracted extensive research interest in the fields of signal processing, image processing, and computer vision \cite{zhang2015survey}.
Specific to communications, sparse optimization was applied for activity detection \cite{chen2018sparse}, random access \cite{liu2018sparse}, and node sleeping \cite{van2020joint}.
Especially, \cite{van2020joint} turned off some ``unnecessary" APs in a CF mMIMO network by formulating a sparse reconstruction problem.
Similarly, in \cite{demir2022cell}, binary programming is utilized to activate only the minimal subsets of APs for each UE to minimize the end-to-end network power consumption.
Inspired by these approaches, we formulate a new sparse optimization problem that minimizes the mean-squared-error (MSE), to enforce sparsity on the LSFD coefficients.
The major contributions are as follows:\vspace{-0.2em}
%Our major contributions are summarized as follows:
\begin{itemize}
  \item We propose the new sparse LSFD (S-LSFD) design where joint AP-UE association and LSFD is achieved by formulating a sparsity-inducing MSE-minimizing problem to force small LSFD coefficients to zero.
  \item We solve the sparsity problem by developing a proximal algorithm with block-coordinate descent (BCD) that is tailored to our problem.
  \item We compare the proposed S-LSFD, the optimal LSFD (O-LSFD) \cite{nayebi2016performance}, and the partial LSFD (P-LSFD) \cite{nayebi2016performance} with the separate association algorithm in \cite{bjornson2020scalable}, considering the SE and EE and different combining schemes.
\end{itemize}

{\bf {\emph {Notation}}}:
Boldface lowercase letters, $\bf x$, denote column vectors, boldface uppercase letters, $\bf X$, denote matrices, and calligraphic uppercase letters, $\cal A$, denote sets.
The superscripts $^{\Ttran}$, $^\star$, and $^{\Htran}$ denote the transpose, conjugate, and conjugate transpose, respectively.
${\bf I }_n$ denotes the $n \!\times \! n$ identity matrix, ${\mathbb E}\{ \cdot \}$ denotes the expected value, and ${\cal N}_{\mathbb C}\left({{\bf 0},{\bf R}}\right)$ denotes the multi-variate circularly symmetric complex Gaussian distribution.
% with correlation matrix $\bf R$.
$x_i \!=\! [{\bf x}]_i$, $(x)_+ \!=\! \max(x,0)$, and ${\rm{sign}}(\cdot)$ is the signum function. %where ${\rm{sign}}(x) = 1 $ if $x>0$, ${\rm{sign}}(x) = 0 $ if $x=0$, and ${\rm{sign}}(x) = -1 $ if $x<0$.

\section{CF mMIMO System Model}

We consider a CF mMIMO system consisting of $K$ single-antenna UEs and $L$ geographically distributed APs, each equipped with $N$ antennas.
As illustrated in Fig.~\ref{fig:system}, we adopt the user-centric CF architecture, where each UE is served by a subset of APs based on the UE-dependent channel quality
and service requirement.
All APs are connected via fronthaul connections to a CPU, which is responsible for coordinating and processing the signals of all UEs.
The AP-UE association is optimized in Section \ref{sec:S-LSFD}.
For now, we let ${\cal D}_l \!\subset\!\{{ 1,\ldots, K }\}$ denote the subset of UEs served by AP $l$ and ${\cal M}_k \!\subset \!\{{ 1,\ldots,L }\}$ denote the subset of APs serving UE $k$.\vspace{-0.7em}
%The procedure for the AP-UE association is elaborated in Section \ref{sec:proximal}.
%For now, we let ${\cal M}_k \subset \left\{{ 1,\ldots,L }\right\}$ denote the subset of APs associated with UE $k$ and let $ {\bf D}_{kl}= {\bf I}_{N}$ if $l \in {\cal M}_k$ and ${\bf D}_{kl}= {\bf 0}_{N}$ otherwise, $\forall k,l$.

\subsection{Channel Model}
We denote by ${\bf h}_{kl} \in {\mathbb C}^N$ the channel between AP $l$ and UE $k$ and assume that these channels are constant in time-frequency coherence blocks of $\tau_c$ channel uses \cite{bjornson2017massive}.
We assume spatially correlated Rayleigh fading, which implies that in each block ${\bf h}_{kl}$ takes an independent realization  according to the distribution\vspace{-0.5em}
\begin{equation}
  {{\bf{h}}_{kl}} \sim {\cal N}_{\mathbb C} ({\bf 0}, {\bf R}_{kl}),
\end{equation}
where ${\bf{R}}_{kl} \in {\mathbb C}^{N \times N}$ is the spatial correlation matrix, and $\beta_{kl} \buildrel \Delta \over = {\rm tr}({\bf R}_{kl})/N$ is the large-scale fading coefficient that describes pathloss and shadowing.
We assume that AP $l$ knows the correlation matrices $\{{\bf R}_{kl} : k = 1,\ldots,K\}$ to all UEs since the correlation matrices represent the long-term channel statistics, and, thus, can be accurately estimated from the received signals \cite{bjornson2017massive}.
We consider the uplink operation, where each block dedicates $\tau_p$ channel uses for pilots and the remaining $\tau_c -\tau_p$ channel uses for payload data.\vspace{-0.7em}
%Therefore, there are $\tau_p$ mutually orthogonal $\tau_p$-length pilot sequences available in the system.

\subsection{Channel Estimation and Uplink Data Transmission}

During the channel estimation, $\tau_p \!<\! K$ holds in a large network since the coherence block length is limited and thus some UEs have to share pilots if a mutually orthogonal set of pilots is used.
We denote by ${\cal S}_{t_k}$ the set of the {\it co-pilot UEs} sharing pilot $t_k$ with UE $k$.
When the UEs in ${\cal S}_{t_k}$ transmit pilot $t_k$, the pilot signal ${\bf{y}}_{{t_k}l}^{\rm{p}} \in {{\mathbb C}^N}$ received at AP $l$ (after taking the inner product of the received signal and the pilot $t_k$) is \cite[Sec. 3]{bjornson2017massive}\vspace{-0.5em}
\begin{equation}\label{eq:received pilot signal at AP}
  {\bf y}_{{t_k}l}^{\rm{p}} = \sum\nolimits_{i \in {{\cal S}_{t_k}}} {\sqrt {{\tau _p}{\rho _{\rm{p}}}} {{\bf{h}}_{il}}}  + {{\bf{n}}_{{t_k}l}},
\end{equation}
where $\rho_{\rm p}$ denotes the pilot transmit power and ${{\bf{n}}_{{t_k}l}} \!\sim\! {{\cal N}_{\mathbb{C}}}\left( {{\bf{0}},{\sigma ^2}{{\bf{I}}_N}} \right)$ is the receiver noise.
The \emph{minimum mean-squared-error (MMSE)} estimate of ${\bf h}_{kl}$ is \cite[Sec. 3]{bjornson2017massive}
\begin{equation}\label{eq:mmse estimate}
  {{{\bf{\widehat h}}}_{kl}} = \sqrt {{\tau _p}{\rho _{\rm{p}}}} {{\bf{R}}_{kl}}{\bf{\Psi }}_{{t_k}l}^{ - 1}{\bf y}_{{t_k}l}^{\rm{p}} \sim {\cal N}_{\mathbb C}\left({ {\bf 0},{\bf B}_{kl} }\right),
\end{equation}
where ${{\bf{\Psi }}_{{t_k}l}} = \sum\nolimits_{i \in {{\cal S}_{t_k}}}  {{\tau _p}{\rho _{\rm{p}}}{{\bf{R}}_{il}}}  + {\sigma ^2}{{\bf{I}}_N}$
is the correlation matrix of ${\bf y}_{{t_k}l}^{\rm{p}}$ in \eqref{eq:received pilot signal at AP} and ${\bf B}_{kl} = {\tau _{\rm{p}}}{\rho _{\rm{p}}}{{\bf{R}}_{kl}}{\bf{\Psi }}_{{t_k}l}^{ - 1}{{\bf{R}}_{kl}}$.

During the uplink data transmission, all APs will receive a superposition of the signals sent from all UEs. The received signal ${\bf y}_{l}^{\rm{ul}} \in \mathbb {C}^{N}$ at AP $l$ is
\begin{equation}\label{eq:}
	{\bf y}_{l}^{\rm{ul}} = \sum\nolimits_{i=1}^{K} {\bf h}_{il}s_{i} + {\bf n}_{l}
\end{equation}
where $s_{i} \in \mathbb{C}$ is the signal transmitted from UE $i$ with the power $p_i = \mathbb{E} \{ |s_i|^2 \}$ and ${\bf n}_{l}  \sim \CN({\bf 0},\sigma^2{\bf I}_{N})$ is the independent additive receiver noise.

Each AP $l$ selects the local combining vector ${\bf v}_{kl}$ for UE $k$ and then computes its local estimate of $s_k$ as
$
\widehat s_{kl} = \vect{v}_{kl}^{\Htran}  {\bf y}_{l}^{\rm{ul}}
$.
One choice with interference suppressing capability is the {\it Local MMSE (L-MMSE)} combining \cite{bjornson2020scalable} given as\vspace{-0.5em}
\begin{equation} \label{eq:lmmse combining}
{\bf v}_{kl} \!=\! p_k \left( \sum\nolimits_{i=1}^K p_i \left({\widehat{\bf h}}_{il} {\widehat{\bf h}}_{il}^{\Htran} \!+\! {\bf R}_{il} \!-\! {\bf B}_{il}\right) + \sigma^2 {\bf I}_N \right)^{-1}{\widehat{\bf h}}_{kl}\vspace{-0.5em}
\end{equation}
which minimizes the local MSE ${\mathbb E}\{ |s_k - {\widehat s}_{kl}|^2| \{{\widehat{\bf h}_{il}}\}  \}$.
Alternatively, maximum ratio (MR) processing scheme with ${\bf v}_{kl} = {\widehat{\bf h}}_{kl}$ can be used.
Note that, for UE $k$, although ${\bf v}_{kl} \ne {\bf 0}$ for all APs, only the APs in ${\cal M}_k$ need to compute ${\bf v}_{kl}$.

Next, all APs deliver the local data estimates to the CPU, which computes its final estimate of $s_k$ as a linear combination of the local estimates:\vspace{-0.25em}
\begin{equation} \label{eq:global data etimate}
\widehat{s}_k = \sum\nolimits_{l=1}^{L} a_{kl}^{\star} \widehat{s}_{kl} = \sum\nolimits_{l=1}^{L} a_{kl}^{\star} \vect{v}_{kl}^{\Htran}  {\bf y}_{l}^{\rm{ul}},\vspace{-0.25em}
\end{equation}
where $a_{kl} \in \mathbb{C}$ is the weight that the CPU assigns to the local signal estimate that AP $l$ has of the signal from UE $k$.
%To limit the fronthaul signaling, the APs are only sending the local data estimates to the CPU and not the channel estimates. Hence, the CPU needs to select the weights  $\{ a_{kl} : k=1, \ldots, K \}$ as a deterministic function of the channel statistics.
In LSFD, the CPU selects the weights  $\{ a_{kl} \}$ as a deterministic function of the channel statistics (to avoid passing around channel estimates). It was introduced for CF mMIMO in \cite{nayebi2016performance} but was originally proposed for Cellular mMIMO \cite{adhikary2017uplink}.
Note that the user-centric architecture is determined by which variables in $\{a_{kl}\}$ are non-zero, so that ${\cal M}_k = \{l:a_{kl} \ne 0\}$ for UE $k$.

By letting ${\bf a}_k = [a_{k1},\ldots,a_{kL}]^{\Ttran} \in {\mathbb C}^L$ denote the LSFD vector of UE $k$ and ${\bf g}_{ki} = [{\bf v}_{k1}^{\Htran}{\bf h}_{i1},  \ldots, {\bf v}_{kL}^{\Htran}{\bf h}_{iL}]^{\Ttran} \in {\mathbb C}^L$ denote the receive-combined channels between UE $i$ and all APs, we rewrite the estimate of $s_k$ in \eqref{eq:global data etimate} as\vspace{-0.5em}
\begin{equation}\label{eq:s etimate}
\widehat{s}_k = {\bf a}_{k}^{\Htran}{\bf g}_{kk}s_{k} + \sum\nolimits_{i=1, i\ne k}^{K} {\bf a}_{k}^{\Htran}{\bf g}_{ki}s_{i} + n'_k,\vspace{-0.5em}
\end{equation}
where $n^\prime_{k} = \sum_{l=1}^{L} a_{kl}^{\star} \vect{v}_{kl}^{\Htran}\vect{n}_{l}$ represents the effective noise.
Although the effective channel $\vect{a}_{k}^{\Htran}\vect{g}_{kk}$ is unknown at the CPU, we notice that its average $\mathbb{E}\{\vect{a}_{k}^{\Htran}  \vect{g}_{kk}\} = \vect{a}_{k}^{\Htran} \mathbb{E}\{ \vect{g}_{kk}\}$ is deterministic and non-zero if the receive combining is selected as suggested above.
Therefore, it can be assumed known
and used to compute the following achievable SE.\vspace{-1.0em}

\subsection{Spectral Efficiency and Energy Efficiency}\vspace{-0.25em}

The well-known \emph{hardening bound} \cite[Th. 4.6]{bjornson2017massive} is used to compute an achievable uplink SE at UE $k$ as
\begin{equation}\label{eq:se}
{\sf SE}_k = \left( {1 - {\tau_p}/{\tau_c}} \right){\log _2}\left( {1 + {\sf SINR}_k} \right)
\end{equation}
where the effective signal-to-interference-and-noise ratio (SINR) is given by \cite[Th. 5.4]{cellfreebook}\vspace{-0.25em}
\begin{equation} \label{eq:sinr}
{\sf SINR}_k = \frac{|{\bf a}_k^{\Htran} {\bf b}_k |^2}{{\bf a}_k^{\Htran} {\bf A}_k {\bf a}_k -  |{\bf a}_k^{\Htran} {\bf b}_k |^2} = \frac{|{\bf a}_k^{\Htran} {\bf b}_k |^2} { {\bf a}_k^{\Htran} ( {\bf A}_k - {\bf b}_k {\bf b}_k^{\Htran} ) {\bf a}_k },\vspace{-0.25em}
\end{equation}
with\vspace{-0.5em}
\begin{align}
 {\bf A}_k  &=  \sum\nolimits_{i=1}^K p_i {\mathbb E} \{{\bf g}_{ki} {\bf g}_{ki}^{\Htran}\} \\
\notag        &+ \sigma^2 \diag \left ({\mathbb E} \left\{ \| {\bf v}_{k1} \|^2 \right\}, \ldots, {\mathbb E} \left\{ \| {\bf v}_{kL} \|^2\right \}\right ) \in {\mathbb C}^{L \times L}, \\
{\bf b}_k  &=  \sqrt{p_k} \mathbb{E}\{{\bf g}_{kk}\} \in {\mathbb C}^{L}.
\end{align}\vspace{-1.5em}

Note that the effective SINR in \eqref{eq:sinr} is a generalized Rayleigh quotient with respect to ${\bf a}_k$ and, thus, the O-LSFD vector is \cite[Lemm. B.10]{bjornson2017massive}\vspace{-0.25em}
\begin{equation} \label{eq:O-LSFD}
{\bf a}_k^{\rm opt} = c_k ({\bf A}_k - {\bf b}_k{\bf b}_k^{\Htran} )^{-1} {\bf b}_k
\end{equation}
with $c_k \in {\mathbb C}$ being an arbitrary non-zero scaling coefficient, which leads to the maximum
$
{\sf SINR}_k = {\bf b}_k^{\Htran} ({\bf A}_k - {\bf b}_k{\bf b}_k^{\Htran} )^{-1} {\bf b}_k.
$\vspace{0.25em}

Additionally, the MSE of UE $k$ becomes\vspace{-0.25em}
\begin{equation}\label{eq:mse}
    {\sf MSE}_k = \mathbb{E}\{ |s_k-\widehat{s}_k |^2 \} = {\bf a}_k^{\Htran} {\bf A}_k {\bf a}_k - 2 \sqrt{p_k} \Re ({\bf a}_k^{\Htran} {\bf b}_k)+p_k,\vspace{-0.25em}
\end{equation}
which is minimized by the O-LSFD vector in \eqref{eq:O-LSFD} with $c_k = \sqrt{p_k} \left(1-{\bf b}_k^{\Htran}{\bf A}_k^{-1}{\bf b}_k\right)$.
Hence, we can conclude that the O-LSFD vector ${\bf a}_k^{\rm opt}$ also minimizes ${\sf MSE}_k$ as \vspace{-0.5em}
\begin{equation} \label{eq:lsfd min mse}
{\bf a}_k^{\rm opt} = \argmin{{\bf a}_k \in {\mathbb C}^{L}} \ {\sf MSE}_k.
\end{equation}

The uplink EE (in bit/Joule) is defined as \cite{bjornson2017massive,ngo2017total}
\begin{equation}\label{eq:ee}
 {\sf{EE}} = \frac{B \cdot \sum\nolimits_{k = 1}^K {{\sf{SE}}_k}}{ \sum\nolimits_{k = 1}^K P_k^{\rm ue} + \sum\nolimits_{l = 1}^L P_l^{\rm ap} + \sum\nolimits_{l = 1}^L P_l^{\rm fh} + P_{\rm cpu}}
\end{equation}
where $B$ is the system bandwidth. 
In the denominator, \vspace{-0.5em}
\begin{equation}\label{eq:}
 P_k^{\rm ue} = P_k^{\rm c,ue} + p_k / \eta_k
\end{equation}
is the power consumption related to UE $k$ where $0 < \eta_k \le 1$ is the power amplifier efficiency and $P_k^{\rm c,ue}$ is the internal circuit power. Moreover,\vspace{-0.5em}
\begin{equation}\label{eq:}
P_l^{\rm ap} = N P_l^{\rm c,ap} + N |{\cal D}_l|\cdot P_l^{\rm proc}
\end{equation}
includes the circuit power $P_l^{\rm c,ap}$ per AP antenna and the power $P_l^{\rm proc}$ for processing the received signal of each served UE at AP $l$. The power related to the fronthaul connections is
\begin{equation}\label{eq:}
 P_l^{\rm fh} = P_l^{\rm fix} + |{\cal D}_l|\cdot P_l^{\rm sig},
\end{equation}
where $P_l^{\rm fix}$ is the fixed power consumption and $P_l^{\rm sig}$ is proportional to the UE-related signaling. Finally,
\begin{equation}\label{eq:}
 P_{\rm cpu} = P_{\rm cpu}^{\rm fix} + \sum\nolimits_{k =1}^{K } \!|{\cal M}_k| \cdot  P_{\rm cpu}^{\rm lsfd} + B \sum\nolimits_{k =1}^{K } \!{{\sf{SE}}_k} \cdot P_{\rm cpu}^{\rm deco}
\end{equation}
where $P_{\rm cpu}^{\rm fix}$ is the fixed power consumption, $P_{\rm cpu}^{\rm lsfd}$ is the power consumption required for LSFD, and $P_{\rm cpu}^{\rm deco}$ is the traffic-dependent power for the final decoding at the CPU.
Typical values for these parameters are given later in Table \ref{tab:paremeter}.\vspace{-0.75em}

\section{Problem Formulation and Sparse LSFD}\label{sec:S-LSFD}

Recall from \eqref{eq:lsfd min mse} that ${\bf a}_k^{\rm opt}$ minimizes ${\sf MSE}_k$. This vector typically contains a few large and many small values, due to pathloss differences between APs and UE. We will develop a sparse LSFD design that put such values zero.
By letting ${\bf{a}} \!=\! {[ {{\bf{a}}_1^{\Ttran}, \!\ldots\! ,{\bf{a}}_K^{\Ttran}} ]^{\Ttran}}\in \mathbb{C}^{KL}$, ${\bf{b}} \!=\! {[ \sqrt{p_1}{{\bf{b}}_1^{\Ttran}, \!\ldots\! ,\sqrt{p_K} {\bf{b}}_K^{\Ttran}} ]^{\Ttran}}\in \mathbb{C}^{KL}$, and ${\bf{A}} \!=\! {\diag}( {{{\bf{A}}_1}, \!\ldots\! ,{{\bf{A}}_K}} )\in \mathbb{C}^{KL\times KL}$, we can cast a similar sparse optimization problem in a generic form as
\begin{equation}\label{eq:p0}
 \min_{{\bf a} \in {\mathbb C}^{KL}}
 {\bf a}^{\Htran} {\bf A}{\bf a} - 2\Re ( {\bf a}^{\Htran}{\bf b} ) + \Omega ( {\bf a} )
\end{equation}
where $\Omega ( {\bf a} )$ is a penalty function that can be designed to turn small values in ${\bf a }_k^{\rm opt}$ into zero.
We propose to use the sparsity-inducing penalty function\vspace{-0.5em}
\begin{equation}\label{eq:}
\Omega({\bf a}) = \gamma \sum\nolimits_{l=1}^L \| {\pmb \alpha}_{l}\|_2 + \lambda \| {\bf a} \|_1
\end{equation}
where ${\pmb \alpha}_l = [a_{1l},\ldots,a_{Kl}]^{\Ttran} \in {\mathbb C}^K$ is the LSFD vector related to AP $l$, and $\gamma$ and $\lambda$ are tunable regularization parameters. 
Lager value of $\gamma$ or $\lambda$ induces more sparsity on vector $\bf a$.
The first term in $\Omega({\bf a})$ is a $\ell_1/\ell_2$-norm regularization that behaves like a $\ell_1$-norm applied to the vector $[\| {\pmb \alpha}_{1}\|_2,\ldots,\| {\pmb \alpha}_{L}\|_2 ]^{\Ttran}$. Element $l$ is small if AP $l$ has little impact on the decoding and the regularization promotes making such values identically zero (i.e., inactivate the AP).
The second term limits the number of the UEs served by the remaining active APs.
We notice that \eqref{eq:p0} is convex since the ``MSE" cost ${\bf a}^{\Htran} {\bf A}{\bf a} - 2\Re ( {\bf a}^{\Htran}{\bf b} )$ and the composite norm penalty $\Omega({\bf a})$ are both convex functions.

Since \eqref{eq:p0} is convex with a non-smooth sparsity-inducing penalty and separable between the APs, proximal methods with BCD can be utilized to reach the global optimum of the problem \eqref{eq:p0} \cite{rish2014sparse}.
We refer to the minimizers of \eqref{eq:p0} as the S-LSFD vectors for $\{ {\widehat s}_k \}$ in \eqref{eq:global data etimate}.\vspace{-0.75em}

\subsection{Proximal Algorithm with Block-Coordinate Descent}

The BCD approach requires \eqref{eq:p0} to be separated into $L$ groups, each related to an AP.
To this end, we first equivalently rewrite the original problem in \eqref{eq:p0} as
\begin{align}\label{eq:p2.1}
 \min_{{\bf a} \in {\mathbb C}^{KL}}
  \left \| {\bar{\bf b}} - \sum\nolimits_{l=1}^L {\bar{\bf A}}_l {\pmb \alpha}_l \right\|_2^2 + \Omega({\bf a}),
\end{align}
where ${\bar{\bf A}}^{\Htran} {\bar{\bf A}} = {\bf A}$ and ${\bar{\bf A}}_l$ is the submatrix of ${\bar{\bf A}}$ with columns corresponding to group $l$ so that we have ${\bar{\bf A}}{\bf a}=\sum_{l=1}^L{\bar{\bf A}}_l{\pmb \alpha}_l$.
The vector ${\bar{\bf b}}$ is defined as ${\bar{\bf b}} = ({\bar{\bf A}}^{\Htran})^{-1}$ {\bf b}.
\eqref{eq:p2.1} has the same form as the so-called ``{\it sparse-group Lasso}" problem \cite{simon2013sparse} but with complex variables.
Then, by using the BCD approach, we can solve \eqref{eq:p2.1} by iteratively minimizing a subproblem of group $l$ while fixing the coefficients of the other groups:
\begin{equation}\label{eq:p2.2}
 {\sf P}_l: \quad
 \min_{ {\pmb \alpha}_l \in {\mathbb C}^{K} }
  { f}({\pmb \alpha}_l) + \Omega'({\pmb \alpha}_l),\quad l=1,\dots,L,
\end{equation}
where ${ f}({\pmb \alpha}_l) = \| {\bf r}_l - {\bar{\bf A}}_l {\pmb \alpha}_l \|_2^2$, $\Omega'({\pmb \alpha}_l) = \gamma \| {\pmb \alpha}_{l}\|_2 + \lambda \| {\pmb \alpha}_{l} \|_1$, and ${\bf r}_l = {\bar{\bf b}} - \sum_{j\ne l} {\bar{\bf A}}_j {\pmb \alpha}_j$ is the partial residual of ${\bar{\bf b}}$ subtracting all group coefficients other than group $l$.
$\Omega'({\pmb \alpha}_l)$ implies that for group $l$, the other group coefficients are considered to be fixed and thus the penalties corresponding to the coefficients in these groups are ignored.
Since ${\bf a}_k$, ${\bf A}_k$, and ${\bf b}_k$ have complex entries, we employ the standard transformations
\begin{equation}\label{eq:comp2real1}
 {\underline{\pmb \alpha}}_l \!=\! \begin{bmatrix} \Re( {\pmb \alpha}_l ) \\ \Im( {\pmb \alpha}_l ) \end{bmatrix}\!,\
 {\underline{\bf A}}_l \!=\! \begin{bmatrix}  \Re( {\bar{\bf A}}_l ) \!\!&\!\!-\Im( {\bar{\bf A}}_l ) \\ \Im( {\bar{\bf A}}_l ) \!\!&\!\!\Re( {\bar{\bf A}}_l ) \end{bmatrix}\!,\
 {\underline{\bf r}}_l \!=\! \begin{bmatrix} \Re( {\bf r}_l ) \\ \Im( {\bf r}_l ) \end{bmatrix},
\end{equation}
and rewrite ${\sf P}_l$ in the following real form:
\begin{equation}\label{eq:p2.3}
 {\bar{\sf P}}_l:\quad
 \min_{ {\underline{\pmb \alpha}}_l \in {\mathbb R}^{2K} }
  {f'}({\underline{\pmb \alpha}}_l) + \Omega{'}({\underline{\pmb \alpha}}_l),
\end{equation}
where ${f'}({\underline{\pmb \alpha}}_l) = \| {\underline{\bf r}}_l - {\underline{\bf A}}_l {\underline{\pmb \alpha}}_l \|_2^2$.
The original ${\sf P}_l$ is upper bounded by its real form ${\bar{\sf P}}_l$ since $\| {\pmb \alpha}_{l} \|_1 \le \| {\underline{\pmb \alpha}}_l \|_1$.

By using proximal methods for a subproblem ${\bar{\sf P}}_l$ of group $l$, we start with an initial point ${\underline{\pmb \alpha}}_l^{0}$ and then compute a sequence of updates ${\underline{\pmb \alpha}}_l^{n}$ that converges to the solution of \eqref{eq:p2.3}, where $n$ is the iteration index.
Given the current ${\underline{\pmb \alpha}}_l^{n}$ obtained at iteration $n$, the next update ${\underline{\pmb \alpha}}_l^{n+1}$ is found by minimizing the following {\it proximal problem}
\begin{equation}\label{eq:p2.4}
\min_{ {\underline{\pmb \alpha}}_l \in {\mathbb R}^{2K} }
 \frac{1}{2} \left\| {\underline{\pmb \alpha}}_l - G( {\underline{\pmb \alpha}}_l^{n}) \right \|_2^2 + \mu \Omega{'}( {\underline{\pmb \alpha}}_l ) ,
\end{equation}
where $G( {\underline{\pmb \alpha}}_l^{n}) = {{\underline{\pmb \alpha}}_l^{n} - \mu \nabla{f'} ({\underline{\pmb \alpha}}_l^{n} )}$ is the so-called {\it gradient update} and $\mu$ is the step length, which is generally computed in practice via line search.
Due to the strong convexity, the unique solution of \eqref{eq:p2.4} can be found \cite{rish2014sparse}.
\begin{lemma}\label{lemm:prox}
Since $\nabla{ f'} ({\underline{\pmb \alpha}}_l ) = 2{\underline{\bf A}}_l^{\Ttran}( {\underline{\bf A}}_l {\underline{\pmb \alpha}}_l -  {\underline{\bf r}}_l )$, the unique solution of \eqref{eq:p2.4} can be computed in closed form as\vspace{-0.25em}
\begin{equation}\label{eq:prox_l1+l2}
{\rm Prox}_{\mu,\Omega{'}} ( G( {\underline{\pmb \alpha}}_l^{n}) ) = {\rm Prox}_{\mu \gamma,\ell_2} \circ {\rm Prox}_{\mu \lambda,\ell_1} ( G( {\underline{\pmb \alpha}}_l^{n}) ),\vspace{-0.25em}
\end{equation}
where $f \circ g (x) \buildrel \Delta \over = f(g(x))$ for any function $f$ and $g$,\vspace{-0.25em}
\begin{equation}\label{eq:prox l1}
[ {\rm Prox}_{\mu,\ell_1} ({\bf x}) ]_i = {\rm{sign}}( x_i ) \cdot { ( {| x_i | - \mu } )_+ }\vspace{-0.25em}
\end{equation}
is the proximal operator of the $\ell_1$-norm \cite{rish2014sparse}, and\vspace{-0.25em}
\begin{equation}\label{eq:prox l2}
{\rm Prox}_{\mu,\ell_2} ({\bf x}) = \begin{cases}
{\frac{\bf x}{\|{\bf x}\|_2}{ ( \|{\bf x}\|_2 - \mu } )_+,}&{{\rm if}\ {\bf x} \ne {\bf 0},}\\
{{\bf 0},}&{{\rm otherwise},}
\end{cases}\vspace{-0.25em}
\end{equation}
is the proximal operator of the $\ell_2$-norm \cite{rish2014sparse}.
\end{lemma}\vspace{-0.5em}
\begin{IEEEproof}
The proof follows a similar approach as in \cite{simon2013sparse} but for problem \eqref{eq:p2.1} and is omitted due to limited space.
\end{IEEEproof}

The minimizer of group $l$ is updated as\vspace{-0.5em}
\begin{equation}\label{eq:}
{\underline{\pmb \alpha}}_l^{n+1} \leftarrow {\rm Prox}_{\mu,\Omega{'}} ( G( {\hat{\underline{\pmb \alpha}}}_l^{n} ) )
\end{equation}
with the Nesterov step ${\hat{\underline{\pmb \alpha}}}_l^{n} = {\underline{\pmb \alpha}}_l^{n} + \frac{n-1}{n+2} ({\underline{\pmb \alpha}}_l^{n} - {\underline{\pmb \alpha}}_l^{n-1})$ which accelerates the convergence \cite{rish2014sparse}, and then is fixed while the other groups are minimized until next iteration.
By iteratively updating $\{ {\underline{\pmb \alpha}}_l: l=1,\ldots,L \}$, the global solution for $\{ {\bar{\sf P}}_l : l=1,\ldots,L\}$ in \eqref{eq:p2.3} can be reached.
Then with the inverse transformation of \eqref{eq:comp2real1} and some algebra, we achieve the complex S-LSFD vectors $\{ {\bf a}_k^{\rm s}: k=1,\ldots,K \}$.\vspace{-1.0em}

\subsection{Sparse AP-UE Association and Power Control}\label{sec:}

According to the S-LSFD vectors $\{ {\bf a}_k^{\rm s}: k=1,\ldots,K \}$, AP $l$ does not serve UE $k$ if $a_{kl}^{\rm s} = 0$ since AP $l$ will not contribute to the global estimate of $s_k$ in \eqref{eq:global data etimate}.
Therefore, we have ${\cal M}_k =\{ l:{a}_{kl}^{\rm s} \ne 0 \}$ for UE $k$.
In the next section, we adopt the fractional power control \cite{chen2020structured}\vspace{-0.5em}
\begin{equation}\label{eq:power control}
{p_k} = \frac{ \min_{i\in \{1,\ldots,K\}}\left(\sum\nolimits_{l\in {\cal M}_i} \beta_{il}\right)^\theta }{\left(\sum\nolimits_{l\in {\cal M}_k} \beta_{kl}\right)^\theta} p_{\rm max},
\end{equation}
where $p_{\rm max}$ is the maximal uplink transmit power of a UE and $\theta \in \left[{0,1}\right]$ determines the power control behavior.
During the optimization, \eqref{eq:power control} is used with ${\cal M}_k = \{1,\ldots,L\},\ \forall k$.\vspace{-1em}

\section{Numerical Results}\label{sec:}\vspace{-0.2em}

In this section, we evaluate the performance of the proposed S-LSFD design in Section \ref{sec:S-LSFD} and validate the algorithm convergence by comparing it to CVX \cite{cvx2015}.
We consider a setup with $K = 20$ UEs that are independently and uniformly dropped in a $0.5\times 0.5$ km$^2$ area.
The wrap-around technique is used to approximate an infinitely large network.
We use the 3GPP Urban Microcell model to compute the large-scale propagation conditions, such as pathloss and shadow fading.
The SE results with L-MMSE combining are obtained from Monte Carlo simulations while the results with MR combining are analytically computed by \cite[Cor. 2]{bjornson2020scalable}. 
Pilot assignment is performed with the scheme from \cite{bjornson2020scalable}.
The main system parameters are given in Table \ref{tab:paremeter} and originate from \cite{bjornson2019making,ngo2017total}.

\renewcommand\arraystretch{1.3}
\begin{table}[t!]
  \centering
  \fontsize{9}{9}\selectfont
  \caption{System Parameters. }
  \vspace{-0.75em}
  \label{tab:paremeter}
    \begin{tabular}{|p{1.45cm}<{\centering}|p{2.25cm}<{\centering}|p{1.65cm}<{\centering}|p{1.75cm}<{\centering}|}
    \hline
        \bf Parameters  & \bf Values  & \bf Parameters  & \bf Values   \cr\hline
    \hline
        $B$, $\tau_c$, $\tau_p$  & 20\,MHz, 200, 10  & $\eta_k$, $\theta$  & 0.4, 1  \cr\hline
        $\rho_{\rm p}$, $p_{\rm max}$  & 0.1\,W, 0.1\,W  & $P_k^{\rm c,ue}$, $P_l^{\rm c,ap}$ & 0.1\,W, 0.2\,W  \cr\hline
        $P_{\rm cpu}^{\rm fix}$, $P_l^{\rm fix}$  & 5\,W, 0.825\,W  & $P_l^{\rm sig}$  & 0.01\,W \cr\hline
        $P_{\rm cpu}^{\rm deco}$  & 1\,W/(Gbit/s)  &  $P_l^{\rm proc}$, $P_{\rm cpu}^{\rm lsfd}$ & 0.8\,W, 1\,W  \cr\hline
    \end{tabular}
  \vspace{-1em}
\end{table}

\begin{figure}[t!]
\centering
\includegraphics[scale=0.6]{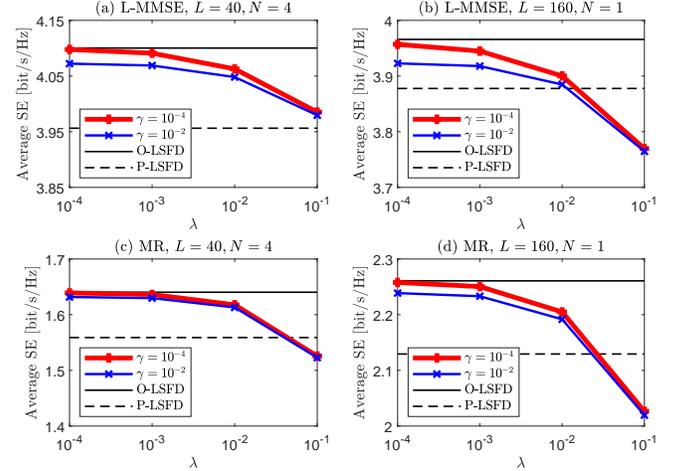}\vspace{-1.5em}
\caption{Average SE with different system setups and combining schemes.
\label{fig:se gw}}
\vspace{-1.75em}
\end{figure}

\begin{figure}[t!]
\centering
\includegraphics[scale=0.6]{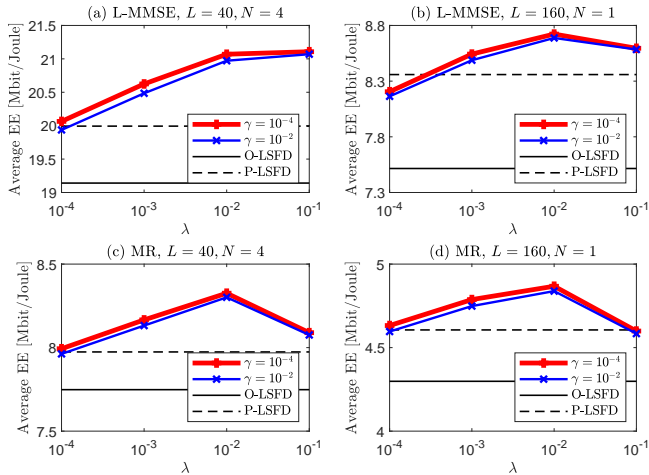}\vspace{-1.5em}
\caption{Average EE with different system setups and combining schemes.
\label{fig:ee gw}}
\vspace{-1.0em}
\end{figure}

We consider two deployment setups: a) $L = 40$, $N = 4$ and b) $L=160$, $N = 1$.
Both setups have the total number of antennas $LN=160$.
To highlight the performance improvements of our joint AP-UE association and LSFD, we consider two benchmarks: the first one is the O-LSFD in \eqref{eq:O-LSFD} where all APs serve all UEs, and the second one applies the heuristic AP-UE association scheme from \cite{bjornson2020scalable} followed by P-LSFD. The latter is referred to as ``P-LSFD" in this section.
In Fig.~\ref{fig:se gw} and Fig.~\ref{fig:ee gw}, we set the vertical scale intervals to emphasize how small/large the gaps are between the curves.

In Fig.~\ref{fig:se gw}, we first evaluate the average SE when the S-LSFD is used and we vary the regularization parameters $\lambda$ and $\gamma$.
The first observation is that both the average SE with L-MMSE combining and MR combining decreases as $\lambda$ increases since the number of  serving APs per UE decreases. These SE results are only {\it slightly} below the O-LSFD but above the P-LSFD (for small values of $\lambda$ that let each UE be served by its most essential APs).
The SE with MR combining drops more markedly than the SE with L-MMSE combining, since the interference suppression makes L-MMSE more robust to a reduction in the number of serving APs.
This feature of L-MMSE also reflects on the SE gaps between the two AP deployment setups, where the L-MMSE case (between Fig.~\ref{fig:se gw}(a) and (b)) has more narrow gaps than the MR case (between Fig.~\ref{fig:se gw}(c) and (d)).
Moreover, we notice that the multi-antenna AP setup outperforms the single-antenna AP setup in the L-MMSE case while it is the opposite in the MR case.
The reason is that in the L-MMSE case, the interference suppression gain enabled by multiple antennas is more beneficial the macro diversity gain brought by having more APs.
In contrast, the macro diversity gain dominates in the MR case, which relies on it for avoiding interference.
When comparing the SE with different $\gamma$, we notice the average SE loss compared to O-LSFD grows when $\gamma$ increases for a similar reason as for $\lambda$.

\begin{figure}[t!]
\centering
\includegraphics[scale=0.6]{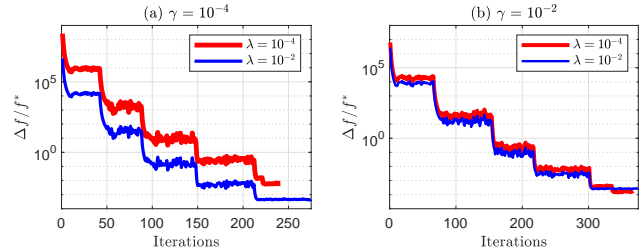}\vspace{-1em}
\caption{Convergence accuracy with different sparsity parameters ($L=40$).
\label{fig:conv}}
\vspace{-1.0em}
\end{figure}

Fig.~\ref{fig:ee gw} shows the average EE of our S-LSFD design.
It is clear that although our S-LSFD slightly reduces the SE, it significantly increases the EE compared to the O-LSFD where all APs serve all UEs. Our joint design also outperform P-LSFD where the AP-UE association and LSFD are performed separately.
We notice that the setup $L=40, N =4$ gives higher EE than the setup $L=160, N =1$ since it consumes more power to have many APs (see \eqref{eq:ee}).
Another observation is that the EE gaps between L-MMSE and MR is larger with multi-antenna APs (between Fig.~\ref{fig:ee gw}(a) and (c)) than with single-antenna APs (between Fig.~\ref{fig:ee gw}(b) and (d)) thanks to the interference suppression.
The average EE is unimodal with respect to $\lambda$ in Fig.~\ref{fig:ee gw}(b), (c), and (d), which implies that there exist a value of $\lambda$ that provides maximum average EE.

The convergence of our proposed proximal algorithm is validated by checking the following two metrics: the accuracy $\Delta f / f^\ast$ and the elapsed time.
The accuracy is defined as the function value difference $\Delta f = f - f^\ast$ normalized by the ``optimal" function value $f^\ast$ obtained by CVX, which is shown in Fig.~\ref{fig:conv}.
The elapsed times for convergence with different sparsity parameters are given in Table \ref{tab:time cost}, where the shadowed results correspond to CVX.
Fig.~\ref{fig:conv} validates the correctness of our algorithm by showing the accuracy of $10^{-4}$.
It can be seen in Fig.~\ref{fig:conv} that the proximal algorithm converges faster with a larger $\lambda$.
The results in Table \ref{tab:time cost} indicate the effectiveness of our algorithm where the elapsed time of our algorithm is much less than that of CVX, especially when $\lambda$ and $\gamma$ are small.

\renewcommand\arraystretch{1.3}
\begin{table}[t!]
  \centering
  \fontsize{9}{9}\selectfont
  \caption{Elapsed time for convergence in Fig.~\ref{fig:conv} (CVX: shadowed). }
  \vspace{-0.75em}
  \label{tab:time cost}
    \begin{tabular}{|p{2.4cm}<{\centering}|p{1cm}<{\centering}|p{1cm}<{\centering}|p{1cm}<{\centering}|p{1cm}<{\centering}|}
    \hline
        Elapsed time [sec]  & \multicolumn{2}{c|}{$\lambda = 10^{-4}$}   & \multicolumn{2}{c|}{$\lambda = 10^{-2}$}  \cr\hline
    \hline
        $\gamma = 10^{-4}$  & \cellcolor{gray!30}152.23  & 82.28  & \cellcolor{gray!30}164.83  & 95.16     \cr\hline
        $\gamma =  10^{-2}$  & \cellcolor{gray!30}166.09  & 120.99  & \cellcolor{gray!30}165.31  & 118.74     \cr\hline
    \end{tabular}
  \vspace{-1.5em}
\end{table}

\section{Conclusion}\label{sec:conclusion}

We proposed a new mechanism for joint AP-UE association and LSFD design for CF mMIMO by formulating a group sparsity problem.
We solved the problem by a proximal algorithm with BCD to intentionally turn off some APs and limit the number of UEs served by each of the remaining active APs.
Numerical results showed that, compared to the O-LSFD and P-LSFD, the proposed S-LSFD gives a better association and more flexibility on the sparsity parameters, and, thus, achieves significant EE improvements with relatively small SE loss, especially when using L-MMSE combining with multi-antenna APs.
Maximum EE could be achieved by tuning $\lambda$.

\bibliographystyle{IEEEtran}
% argument is your BibTeX string definitions and bibliography database(s)
\bibliography{IEEEabrv,ref}

% Generated by IEEEtran.bst, version: 1.13 (2008/09/30)
\begin{thebibliography}{10}
\providecommand{\url}[1]{#1}
\csname url@samestyle\endcsname
\providecommand{\newblock}{\relax}
\providecommand{\bibinfo}[2]{#2}
\providecommand{\BIBentrySTDinterwordspacing}{\spaceskip=0pt\relax}
\providecommand{\BIBentryALTinterwordstretchfactor}{4}
\providecommand{\BIBentryALTinterwordspacing}{\spaceskip=\fontdimen2\font plus
\BIBentryALTinterwordstretchfactor\fontdimen3\font minus
  \fontdimen4\font\relax}
\providecommand{\BIBforeignlanguage}[2]{{%
\expandafter\ifx\csname l@#1\endcsname\relax
\typeout{** WARNING: IEEEtran.bst: No hyphenation pattern has been}%
\typeout{** loaded for the language `#1'. Using the pattern for}%
\typeout{** the default language instead.}%
\else
\language=\csname l@#1\endcsname
\fi
#2}}
\providecommand{\BIBdecl}{\relax}
\BIBdecl

\bibitem{saad2019vision}
W.~Saad, M.~Bennis, and M.~Chen, ``A vision of {6G} wireless systems:
  Applications, trends, technologies, and open research problems,'' \emph{IEEE
  netw.}, vol.~34, no.~3, pp. 134--142, Mar. 2019.

\bibitem{zhang2019multiple}
J.~Zhang, E.~Bj{\"o}rnson, M.~Matthaiou, D.~W.~K. Ng, H.~Yang, and D.~J. Love,
  ``Prospective multiple antenna technologies for beyond {5G},'' \emph{IEEE J.
  Sel. Areas Commun.}, vol.~38, no.~8, pp. 1637--1660, Aug. 2020.

\bibitem{chen2021survey}
S.~Chen, J.~Zhang, J.~Zhang, E.~Bj{\"o}rnson, and B.~Ai, ``A survey on
  user-centric cell-free massive {MIMO} systems,'' \emph{Digit. Commun. Netw.},
  2021.

\bibitem{demir2022cell}
{\"O}.~T. Demir, M.~Masoudi, E.~Bj{\"o}rnson, and C.~Cavdar, ``Cell-free
  massive {MIMO} in virtualized {CRAN}: how to minimize the total network
  power?'' in \emph{Proc. IEEE ICC}, to appear, 2022.

\bibitem{cellfreebook}
{\"O}.~T. Demir, E.~Bj{\"o}rnson, and L.~Sanguinetti, ``Foundations of
  user-centric cell-free massive {MIMO},'' \emph{Foundations and
  Trends{\textregistered} in Signal Processing}, vol.~14, no. 3-4, pp.
  162--472, 2021.

\bibitem{nayebi2016performance}
E.~Nayebi, A.~Ashikhmin, T.~L. Marzetta, and B.~D. Rao, ``Performance of
  cell-free massive {MIMO} systems with {MMSE} and {LSFD} receivers,'' in
  \emph{Proc. 50th Asilomar Conf. Signals, Syst. Comput.}, Nov. 2016, pp.
  203--207.

\bibitem{bjornson2019making}
E.~Bj{\"o}rnson and L.~Sanguinetti, ``Making cell-free massive {MIMO}
  competitive with {MMSE} processing and centralized implementation,''
  \emph{IEEE Trans. Wireless Commun.}, vol.~19, no.~1, pp. 77--90, Jan. 2019.

\bibitem{buzzi2019user}
S.~Buzzi, C.~D'Andrea, A.~Zappone, and C.~D'Elia, ``User-centric {5G} cellular
  networks: Resource allocation and comparison with the cell-free massive
  {MIMO} approach,'' \emph{IEEE Trans. Wireless Commun.}, vol.~19, no.~2, pp.
  1250--1264, Feb. 2019.

\bibitem{bjornson2020scalable}
E.~Bj{\"o}rnson and L.~Sanguinetti, ``Scalable cell-free massive {MIMO}
  systems,'' \emph{IEEE Trans. Commun.}, vol.~68, no.~7, pp. 4247--4261, Jul.
  2020.

\bibitem{zhang2015survey}
Z.~Zhang, Y.~Xu, J.~Yang, X.~Li, and D.~Zhang, ``A survey of sparse
  representation: algorithms and applications,'' \emph{IEEE access}, vol.~3,
  pp. 490--530, 2015.

\bibitem{chen2018sparse}
Z.~Chen, F.~Sohrabi, and W.~Yu, ``Sparse activity detection for massive
  connectivity,'' \emph{IEEE Trans. Signal Process.}, vol.~66, no.~7, pp.
  1890--1904, Jul. 2018.

\bibitem{liu2018sparse}
L.~Liu, E.~G. Larsson, W.~Yu, P.~Popovski, C.~Stefanovic, and E.~De~Carvalho,
  ``Sparse signal processing for grant-free massive connectivity: A future
  paradigm for random access protocols in the internet of things,'' \emph{IEEE
  Signal Process. Mag.}, vol.~35, no.~5, pp. 88--99, 2018.

\bibitem{van2020joint}
T.~Van~Chien, E.~Bj{\"o}rnson, and E.~G. Larsson, ``Joint power allocation and
  load balancing optimization for energy-efficient cell-free massive {MIMO}
  networks,'' \emph{IEEE Trans. Wireless Commun.}, vol.~19, no.~10, pp.
  6798--6812, Oct. 2020.

\bibitem{bjornson2017massive}
E.~Bj{\"o}rnson, J.~Hoydis, and L.~Sanguinetti, ``Massive {MIMO} networks:
  Spectral, energy, and hardware efficiency,'' \emph{Foundations and
  Trends{\textregistered} in Signal Processing}, vol.~11, no. 3-4, pp.
  154--655, 2017.

\bibitem{adhikary2017uplink}
A.~Adhikary, A.~Ashikhmin, and T.~L. Marzetta, ``Uplink interference reduction
  in large-scale antenna systems,'' \emph{IEEE Trans. Commun.}, vol.~65, no.~5,
  pp. 2194--2206, May 2017.

\bibitem{ngo2017total}
H.~Q. Ngo, L.-N. Tran, T.~Q. Duong, M.~Matthaiou, and E.~G. Larsson, ``On the
  total energy efficiency of cell-free massive {MIMO},'' \emph{IEEE Trans.
  Green Commun. Netw.}, vol.~2, no.~1, pp. 25--39, Mar. 2018.

\bibitem{rish2014sparse}
I.~Rish and G.~Grabarnik, \emph{Sparse modeling: theory, algorithms, and
  applications}.\hskip 1em plus 0.5em minus 0.4em\relax CRC press, 2014.

\bibitem{simon2013sparse}
N.~Simon, J.~Friedman, T.~Hastie, and R.~Tibshirani, ``A sparse-group lasso,''
  \emph{J. Comput. Graph. Stat.}, vol.~22, no.~2, pp. 231--245, May 2013.

\bibitem{chen2020structured}
S.~Chen, J.~Zhang, E.~Bj{\"o}rnson, J.~Zhang, and B.~Ai, ``Structured massive
  access for scalable cell-free massive {MIMO} systems,'' \emph{IEEE J. Sel.
  Areas Commun.}, vol.~39, no.~4, pp. 1086--1100, Apr. 2020.

\bibitem{cvx2015}
{CVX Research Inc.}, ``{CVX: Matlab} software for disciplined convex
  programming, academic users,'' {\url{http://cvxr.com/cvx/}}, 2015.

\end{thebibliography}

\end{document}